\documentstyle[pre,aps,floats,epsf]{revtex}

\begin{document}

\draft
 
\twocolumn[\hsize\textwidth\columnwidth\hsize\csname@twocolumnfalse%
\endcsname           

\title{Optimizing Traffic Lights in a Cellular Automaton Model for 
        City Traffic}

\author{Elmar Brockfeld$^{1}$ \and Robert Barlovic$^{2}$ \and Andreas
  Schadschneider$^{3}$ \and Michael Schreckenberg$^{2}$\\}

\address{$^{1}$Deutsches Zentrum f\"ur Luft- und Raumfahrt e.V. (DLR),
51170 K\"oln, Germany,\\ email: Elmar.Brockfeld@dlr.de}

\address{$^{2}$Theoretische Physik FB 10, Gerhard-Mercator-Universit\"at
Duisburg, D-47048 Duisburg, Germany,\\ email: barlovic,schreckenberg@uni-duisburg.de}

\address{$^{3}$ Institut f\"ur Theoretische Physik, Universit\"at zu
K\"oln, D-50937 K\"oln, Germany,\\ email: as@thp.uni-koeln.de}

\date{\today}

\maketitle              

\begin{abstract} 
        
We study the impact of global traffic light control strategies in a
recently proposed cellular automaton model for vehicular traffic in
city networks. The model combines basic ideas of the
Biham-Middleton-Levine model for city traffic and the
Nagel-Schreckenberg model for highway traffic. The city network has a
simple square lattice geometry. All streets and intersections are
treated equally, i.e., there are no dominant streets. Starting
from a simple synchronized strategy we show that the capacity of the
network strongly depends on the cycle times of the traffic
lights. Moreover we point out that the optimal time periods are
determined by the geometric characteristics of the network, i.e., the
distance between the intersections. In the case of synchronized
traffic lights the derivation of the optimal cycle times in the
network can be reduced to a simpler problem, the flow optimization of
a single street with one traffic light operating as a bottleneck. In
order to obtain an enhanced throughput in the model improved global
strategies are tested, e.g., green wave and random switching
strategies, which lead to surprising results. 

\end{abstract}            

\vspace{0.8cm}

]


\section{Introduction}

Mobility is nowadays regarded as one of the most significant
ingredients of a modern society. Unfortunately, the capacity
of the existing street networks is often exceeded. In urban
networks the flow is controlled by traffic lights and
traffic engineers are often forced to question if the capacity of the
network is exploited by the chosen control strategy. One possible
method to answer such questions could be the use of vehicular traffic
models in control systems as well as in the planning and design of
transportation networks. For almost half a century there were strong
attempts to develop a theoretical framework of traffic science. Up to
now, there are two different concepts for modeling vehicular traffic
(for an overview
see~\cite{tgf95,tgf97,tgf99,prig,dag,helbing,helbing01,review}).  In the
``coarse-grained'' fluid-dynamical description, traffic is viewed as a
compressible fluid formed by vehicles which do not appear explicitly
in the theory. In contrast, in the ``microscopic'' models traffic is
treated as a system of interacting particles where attention is
explicitly focused on individual vehicles and the interactions among
them. These models are therefore much better suited for the investigation
of urban traffic.
Most of the ``microscopic'' models developed in recent years
are usually formulated using the language of cellular automata
(CA)~\cite{wolfram}.  Due to the simple nature CA models can be used
very efficiently in various applications with the help of computer
simulations, e.g., large traffic network can be simulated in multiple
realtime on a standard PC.\\

In this paper we analyze the impact of global traffic light control
strategies, in particular synchronized traffic lights, traffic lights
with random offset, and with a defined offset in a recently
proposed CA model for city traffic (see Sec.~\ref{definition} for
further explanation). Chowdhury and
Schadschneider~\cite{ChSch1,ChSch2} combine basic ideas from the
Biham-Middleton-Levine (BML)~\cite{bml1} model of city traffic and the
Nagel-Schreckenberg (NaSch)~\cite{nasch1} model of highway
traffic. This extension of the BML model will be denoted ChSch model
in the following.  \\

The BML model~\cite{bml1} is a simple two-dimensional (square lattice)
CA model. Each cell of the lattice represents a intersection of an
east-bound and a north-bound street. The spatial extension of the
streets between two intersections is completely neglected. The cells
(intersections) can either be empty or occupied by a vehicle moving
to the east or to the north. In order to enable movement in two
different directions, east-bound vehicles are updated at every odd
discrete time-step whereas north-bound vehicles are updated at every
even time-step.  The velocity update of the cars is realized following
the rules of the asymmetric simple exclusion process
(ASEP)~\cite{asep1}: a vehicle moves forward by one cell if the cell
in front is empty, otherwise the vehicle stays at its actual
position. The alternating movement of east-bound and north-bound
vehicles corresponds to a traffic lights cycle of one time-step. In
this simplest version of the BML model lane changes are not possible
and therefore the number of vehicles on each street is
conserved. However, in the last years various modifications and
extensions~\cite{bml2,bml3,bml4,bml5,bml6,bml7} have been proposed for
this model (see also \cite{review} for a review).\\

The NaSch model~\cite{nasch1} is a probabilistic CA model for
one-dimensional highway traffic. It is the simplest known CA model
that can reproduce the basic phenomena encountered in real traffic,
e.g., the occurrence of phantom jams (``jams out of the blue''). In
order to obtain a description of highway traffic on a more detailed
level various modifications to the NaSch model have been proposed and
many CA models were suggested in recent years
(see~\cite{barlo1,knospe1,bril,emm,helbing2}).  The motion in the
NaSch model is implemented by a simple set of rules. The first rule
reflects the tendency to accelerate until the maximum speed
$v_{\text{max}}$ is reached. To avoid accidents, which are forbidden
explicitly in the model, the driver has to brake if the speed exceeds
the free space in front. This braking event is implemented by the
second update rule. In the third update rule a stochastic element is
introduced. This randomizing takes into account the different
behavioral patterns of the individual drivers, especially
nondeterministic acceleration as well as overreaction while slowing
down. Note, that the NaSch model with $v_{\text{max}}=1$ is equivalent
to the ASEP which, in its deterministic limit, is used for the
movement in the BML model.\\
              
One of the main differences between the NaSch model and the BML model
is the nature of jamming. In the NaSch model traffic jams appear
because of the intrinsic stochasticity of the dynamics
\cite{random,nagel}.  The movement of vehicles in the BML model is
completely deterministic and stochasticity arises only from the random
initial conditions.  Additionally, the NaSch model describes vehicle
movement and interaction with sufficiently high detail for most
applications while the vehicle dynamics on streets is completely
neglected in the BML model (except for the effects of hard-core
exclusion). In order to take into account the more detailed dynamics,
the BML model is extended by inserting finite streets between the
cells.  On the streets vehicles drive in accordance to the NaSch
rules. Further, to take into account interactions at the intersections,
some of the prescriptions of the BML model have to be modified. At
this point we want to emphasize that in the considered network all
streets are equal in respect to the processes at intersection, i.e.,
no streets or directions are dominant. The average densities, traffic light
periods etc.\ for all streets (intersections) are assumed to be equal
in the following.\\

The paper is organized as follows: In the next section the definition
of the model
is presented. It will be shown that a simple change of the update
rules is sufficient to avoid the transition to a completely blocked
state that occurs at a finite density in analogy to the BML model
\cite{bml5,bml6,bml7}.  Note, that this blocking is undesirable when
testing different traffic light control strategies and is therefore
avoided in our analyses. In Section~\ref{strategies} different global
traffic light control strategies are presented and their impact on
the traffic will be shown. Further it is illustrated that most of the
numerical results affecting the dependence between the model
parameters and the optimal solutions for the chosen control strategies
can be derived  by simple heuristic arguments in good agreement 
with the numerical results. In the summary we will discuss how the results
can be used benefitably for real urban traffic situations and whether
it could be useful to consider improved control systems, e.g.,
autonomous traffic light control.\\

\section{Definition of the Model}

\label{definition}

\begin{figure}[!hbt]
\begin{center}
\epsfxsize=0.85\columnwidth\epsfbox{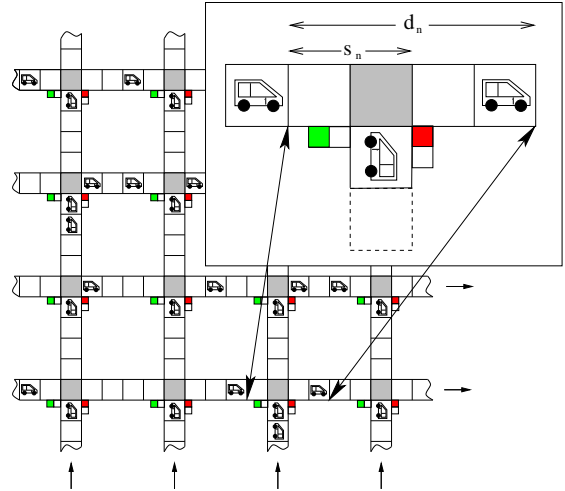}
\end{center}
\vspace{0.2cm}
\caption{Snapshot of the underlying lattice of the model. In this case
the number of intersections in the quadratic network is set to
$N\times N=16$.  The length of the streets between two intersections is
chosen to $D-1=4$. Note that vehicles can only move from west to east
on the horizontal streets or from south to north on the vertical
ones. The magnification on the right side shows a
segment of a west-east street. Obviously the traffic lights are
synchronized and therefore all vehicles moving from south to north
have to wait until they switch to ``green light''.}
\label{model}
\end{figure}

The main aim of the city model proposed in~\cite{ChSch1} is to provide
a more detailed description of city traffic than that of the original
formulation of the BML model. Especially the important interplay of
the different timescales set by the vehicle dynamics, distance between
intersections and cycle times can be studied in the ChSch model.
Therefore each bond of the network is decorated with $D-1$ cells
representing single streets between each pair of successive
intersections. Moreover, the traffic lights are assumed to flip
periodically at regular time intervals $T$ instead of alternating
every time-step ($T>1$).  Each vehicle is able to move forward
independently of the traffic light state, as long as it reaches a site
where the distance to the traffic light ahead is smaller than the
velocity. Then it can keep on moving if the light is green. Otherwise
it has to stop immediately in front of it.

As one can see from Fig.~\ref{model}, the network of streets builds a
$N\times N$ square lattice, i.e., the network consist of $N$
north-bound and $N$ east-bound street segments. The simple square
lattice geometry is determined by the fact that the length of all
$2N^2$ street segments is equal and the streets segments are assumed
to be parallel to the $x-$ and $y-$axis. In addition, all
intersections are assumed to be equitable, i.e., there are no main
roads in the network where the traffic lights have a higher priority.
In accordance with the BML model streets parallel to the $x-$axis
allow only single-lane east-bound traffic while the ones parallel to
the $y-$axis manage the north-bound traffic. The separation between
any two successive intersections on every street consists of $D-1$ cells
so that the total number of cells on every street is $L= N\times
D$. Note, that for $D=1$ the structure of the network corresponds to
the BML model, i.e., there are only intersections without roads connecting
them.\\

The traffic lights are chosen to switch simultaneously after a fixed
time period $T$. Additionally all traffic lights are synchronized,
i.e., they remain green for the east-bound vehicles and they are red
for the north-bound vehicles and vice versa. The length of the time
periods for the green lights does not depend on the direction and thus
the ``green light'' periods are equal to the ``red light'' periods. At
this point it is important to premention that a large part of our
investigations will consider a different traffic light strategy. In
the following the strategy described above will be called
``synchronized strategy''. In addition we improved the traffic lights
by assigning an offset parameter to every one. This modification can
be used for example to shift the switch of two successive traffic
lights in a way that a ``green wave'' can be established in the
complete network. The different ``traffic light strategies'' used here
are discussed in detail in Sec.~\ref{strategies}.\\
      
As in the original BML model periodic boundary conditions are chosen
and the vehicles are not allowed to turn at the intersections.  Hence,
not only the total number $N_v$ of vehicles is conserved, but also the
numbers $N_x$ and $N_y$ of east-bound and north-bound vehicles,
respectively. All these numbers are completely determined by the 
initial conditions. In analogy to the NaSch model
the speed $v$ of the vehicles can take one of the $v_{\text{max}}+1$ integer
values in the range $v=0,1,...,v_{\text{max}}$. The dynamics of vehicles on
the streets is given by the maximum velocity $v_{\text{max}}$ and the
randomization parameter $p$ of the NaSch model which is responsible
for the movement. The state of the network at time $t+1$ can be
obtained from that at time $t$ by applying the following rules to all
cars at the same time (parallel dynamics):\\

\begin{itemize}
\item Step 1: {\it Acceleration:}\\
\phantom{Step 1: }$v_{n}\rightarrow \min(v_{n}+1,v_{{\rm max}})$\\
\item Step 2: {\it Braking due to other vehicles or traffic light state:}
\begin{itemize}
\item Case 1: The traffic light is red in front of the {\it n}-th vehicle:\\
\phantom{Case 1: }$v_{n}\rightarrow \min(v_{n},d_{n}-1,s_{n}-1)$
\item Case 2: The traffic light is green in front of the {\it n}-th vehicle:\\
\phantom{Case 2: }If the next two cells directly behind\\ 
\phantom{Case 2: }the intersection are occupied \\
\phantom{Case 2: else}\quad$v_{n}\rightarrow \min(v_{n},d_{n}-1,s_{n}-1)$  \\
\phantom{Case 2: }else\quad$v_{n}\rightarrow \min(v_{n},d_{n}-1)$\\
\end{itemize}
\item Step 3: {\it Randomization with probability $p$:}\\
\phantom{Step 3: }$v_{n}\rightarrow \max(v_{n}-1,0)$\\
\item Step 4: {\it Movement:}\\
\phantom{Step 4: }$x_{n}\rightarrow x_{n}+v_{n}$
\end{itemize}                         

Here $x_n$ denotes the position of the {\it n-}th car and
$d_n=x_{n+1}-x_n$ the distance to the next car ahead (see Fig.~\ref{model}). 
The distance to the next traffic light ahead is given by $s_n$. 
The length of a single cell is set to $7.5~m$ in accordance 
to the NaSch model.  
The maximal velocity of the cars is set to $v_{\text{max}}=5$ throughout 
this paper.  Since this should correspond to a typical speed limit  of 
$50~km/h$ in cities, one time-step approximately corresponds to $2~sec$ 
in real time. In the initial state of 
the system, $N_v$ vehicles are distributed among the streets. Here we only
consider the case where the number of vehicles on east-bound streets
$N_x=\frac{N_v}{2}$ is equal to the one on north-bound streets
$N_y=\frac{N_v}{2}$. The global density then is defined by
$\rho=\frac{N_v}{N^2(2D-1)}$ since in the initial state the $N^2$ intersections
are left empty.\\ 

Note, that we have modified Case 2 of Step 2 in comparison
to~\cite{ChSch2}. Due to this modification a driver will only be able
to occupy a intersection if it is assured that he can leave it
again. A vehicle is able to leave a intersection if at least the first
cell behind it will become empty. This is possible for most cases
except when the next two cells directly behind the intersection are
occupied.  The modification itself is done to avoid the transition to
a completely blocked state (gridlock) that can occur in the original
formulation of the ChSch model. Further in the original formulation
\cite{ChSch1} the traffic lights mimick effects of a yellow light
phase, i.e., the intersection is blocked for both directions one
second before switching.  This is done to attenuate the transition to
a blocked state (gridlock). Since the blocked states are completely
avoided in our modification we do not consider a yellow light
anymore. The reason for avoiding the gridlock situation in our
considerations is that we focus on the impact of traffic light control on the
network flow, so that a transition to a blocked state would prevent
from exploring higher densities.  Besides relatively small densities
are more relevant for applications to real networks. However, taking
into account that situations where cars are not able to enter an
intersection are extremely rare, it is clear that this modification
does not change the overall dynamics of the model. Moreover we
compared the original formulation of the ChSch model and the modified
one by simulations and found no differences except for the gridlock
situations which appear in the original formulation due to the
stronger interactions between intersections and roads.


\section{Strategies}

\label{strategies}

As mentioned before our main interest is the investigation of global
traffic light strategies. We want to find methods to improve the
overall traffic conditions in the considered model. At this point it
has to be taken into account that all streets are treated as
equivalent in the considered network, i.e., there are no dominant
streets. This makes the optimization much more difficult and implies
that the green and red phases for each direction should have the same
length. For a main road intersection with several minor roads the total flow
usually can be improved easily by optimizing the flow on the main
road.

We first study the dependence between traffic light periods and
aggregated dynamical quantities like flow or mean velocity. It is
shown that investigating the simpler problem of a single road with one
traffic light (i.e., $N=1$) operating as a defect is sufficient to
give appropriate results concerning the overall network behavior. The
results can be used as a guideline to adjust the optimal traffic light
periods in respect to the model and network parameters. Further we
show that a two dimensional green wave strategy can be established in
the whole network giving much improvement in comparison to the
synchronized traffic light switching. Finally we demonstrate that
switching successive traffic lights with a random shift can be very
useful to create a more flexible strategy which does not depend much
on the model and network parameters. Throughout the paper we will
always assume that the duration of green light is equal to the
duration of the red light phase.\\
      
\subsection{Synchronized Traffic Lights}
\label{synchrostrat}

\begin{figure}[!hbt]
\begin{center}
\epsfxsize=0.85\columnwidth\epsfbox{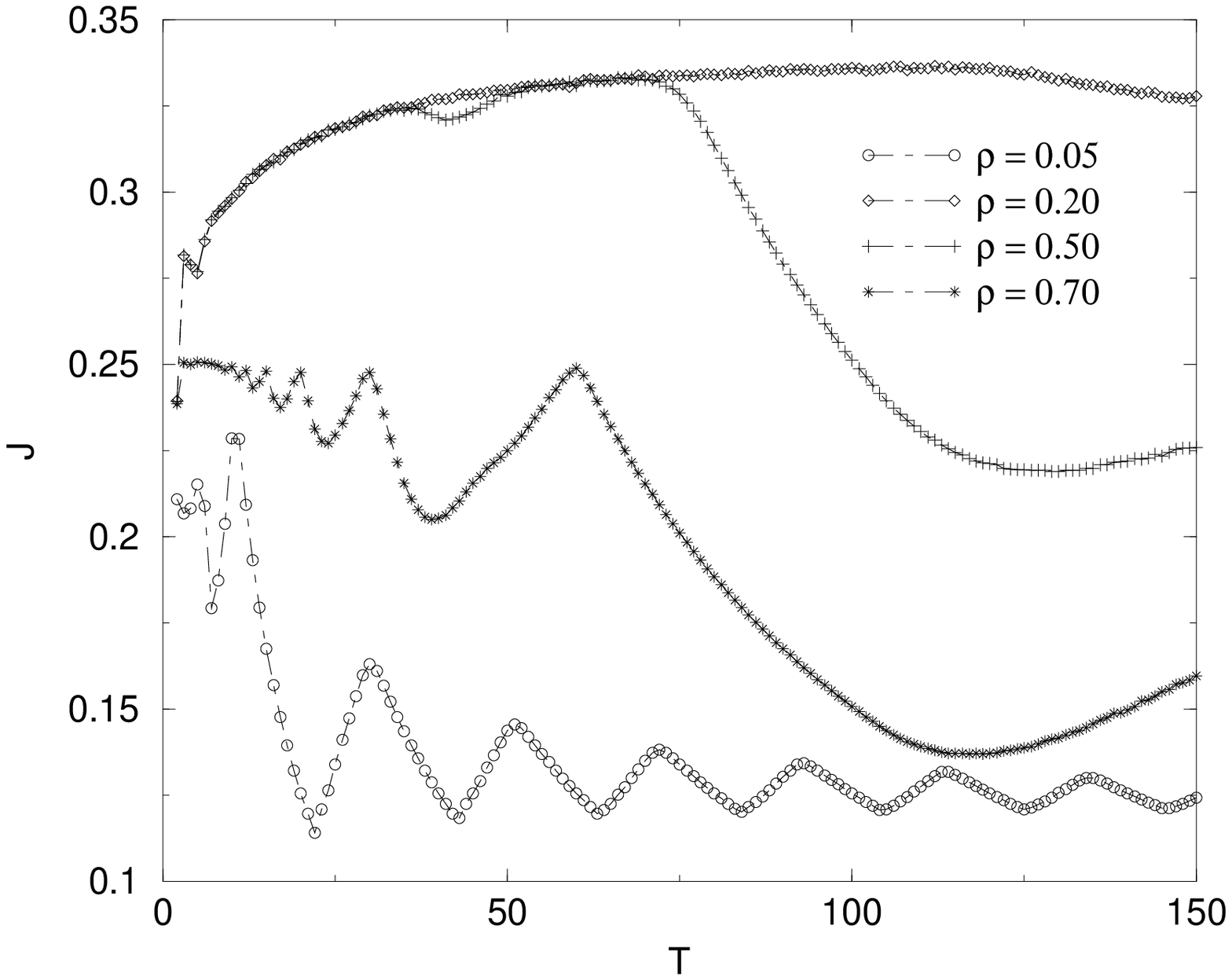}
\epsfxsize=0.85\columnwidth\epsfbox{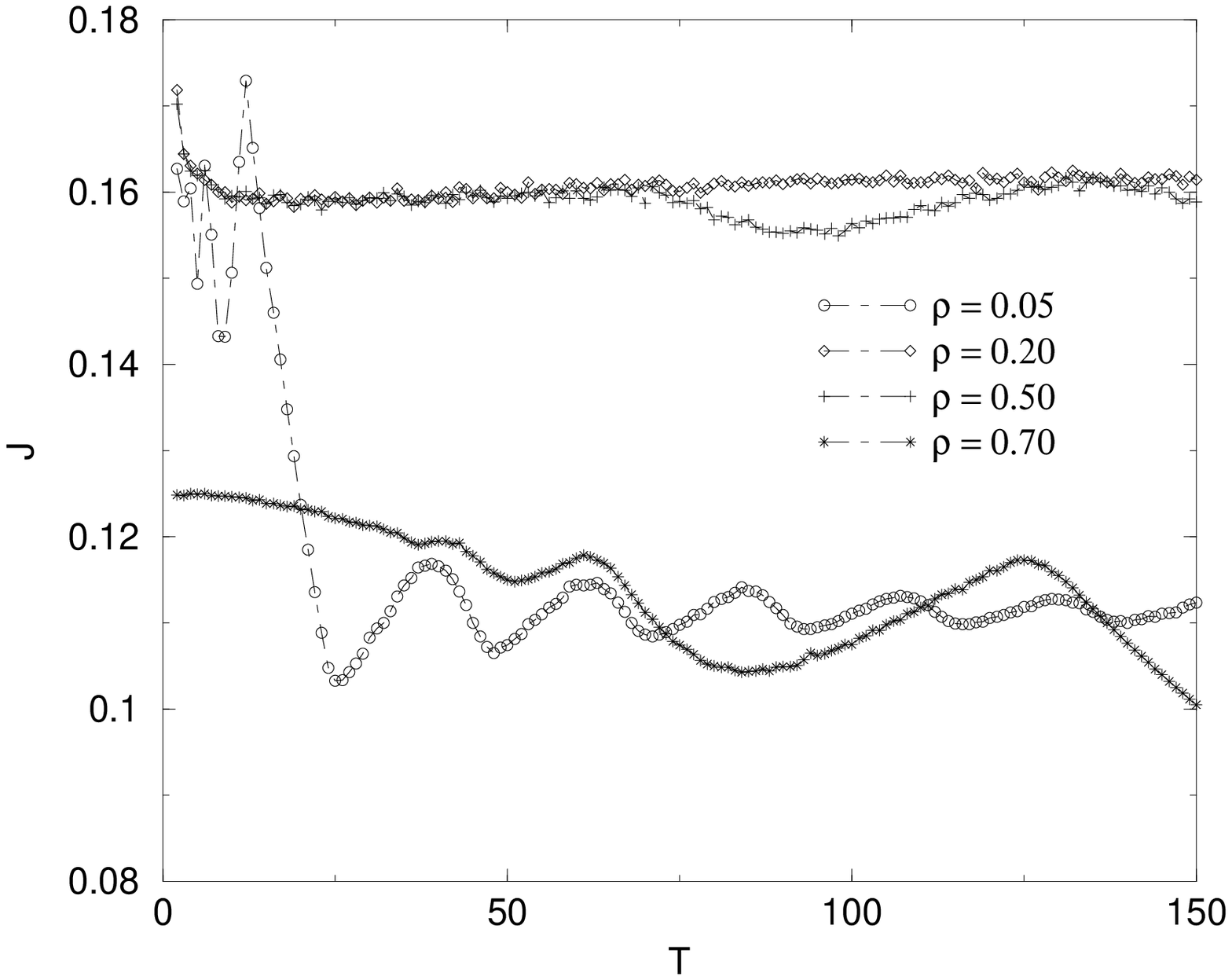}
\end{center}
\vspace{0.2cm}
\caption{The mean flow
$J$ of the smallest network segment (one single intersection, $N=1$) is 
plotted for different global densities as a function of the cycle
length $T$. For the top part of the figure we use a randomization
parameter of $p=0.1$ while in the bottom plot higher fluctuations
$p=0.5$ are considered. In both cases the free-flow regime 
(density $\rho=0.05$) shows a similar shape.  The high
density regime reflects a stronger dependence on the randomization
parameter, but also for the higher $p$ strong variations of the mean
flow can be found. The length of the street is 
$L = 100$ and the flow is aggregated over $100.000$ time-steps.}
\label{jgegenperiod}
\end{figure}   

The starting point of our investigations is the smallest possible
network topology of the ChSch model. Obviously this is a system
consisting of only one east-bound and one north-bound street, i.e.,
$N=1$, linked by a single intersection. As a further simplification we
focus on only one of the two directions of this ``mini'' network,
i.e., a single street with periodic boundary conditions and one
signalized cell in the middle. It is obvious that in the case of one
single traffic light the term ``synchronized'' is a little bit
confusing, but the relevance of this special case to large networks
with synchronized traffic lights will be discussed later.\\
          
Fig.~\ref{jgegenperiod} shows the typical dependence between the
time periods of the traffic lights and the mean flow in the system.
For low densities one finds a strongly oscillating curve with 
maxima and minima at regular distances. In the case of a small
fluctuation parameter $p$ similar oscillations can be even found at very
high densities. For an understanding of the underlying dynamics leading
to such strong variations in the mean flow we take a look into the
microscopic structure. This will allow us to formulate a simple
phenomenological approach which shows a very good agreement with
numerical results. Note that we restrict our investigations to low
densities because for free-flow densities\footnote{Here states are 
denoted as free-flow states if the mean density is smaller
than the density corresponding to the maximum flow of the underlying 
NaSch model.} vehicles are not
constricted by jamming due to the model dynamics, but rather by ``red''
traffic lights. Hence the free-flow density range shows the largest
potential for flow optimization. Later on we will point out the origin
of the oscillating flow even at very high densities which is
completely different to the free-flow case.\\

\begin{figure}[!hbt]
\begin{center}
\epsfxsize=0.85\columnwidth\epsfbox{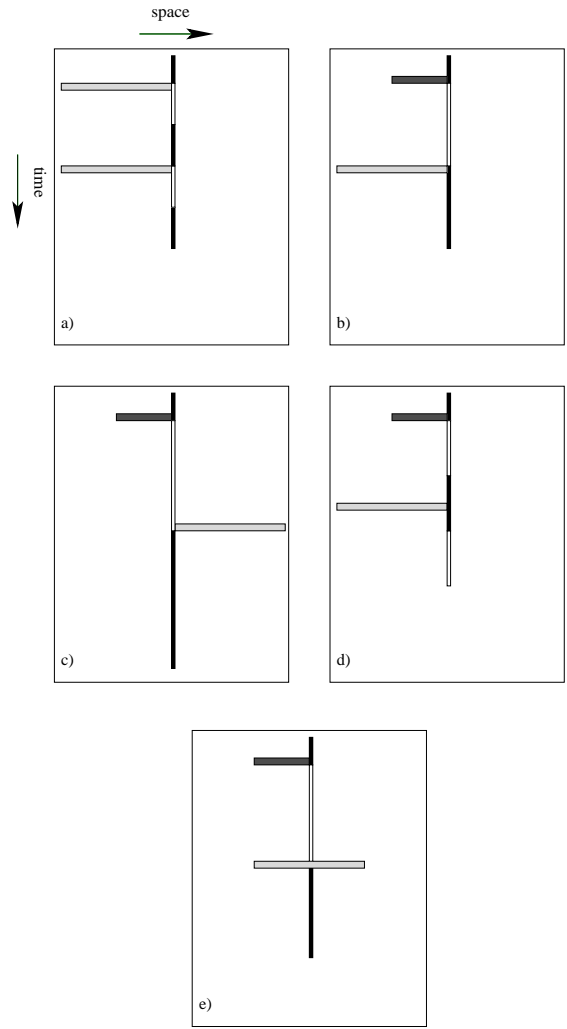}
\end{center}
\vspace{0.2cm}
\caption{Schematic representation of the vehicle movement on
the east bound street for different cycle times.
Standing cars are represented by dark grey rectangles ($x$-axis)
while moving vehicle clusters are bright grey rectangles. The traffic
light is placed in the middle of every figure (time runs along the
$y$-axis). Its state is indicated by the colour of the vertical rectangle.
Green light corresponds to the white colored area of the
traffic light while red light is painted in dark. At this point one
has to take into account that the considered street has periodic
boundary conditions and therefore vehicles leaving the right end of
every scenario (a)--(e) will return after a certain time on the left
side. }
\label{explain2}
\end{figure}             
  
To give an impression of the influence of the cycle times on the
vehicle movement a schematic representation of the observed street is
depicted in Fig.~\ref{explain2}. This picture covers typical dynamical
patterns occurring in the system due to vehicles which are restricted
in their movement by the ``red light''. Based on these scenarios a
simple phenomenological approach is presented in the following which
is able to explain the dependence between vehicle movement and model
parameters. We assume that during one traffic light cycle free-flowing
vehicles form a stable cluster with a width which is approximately
constant. Further we assume that a phase separation between
free-flowing and jammed vehicles takes place at high densities. The
legitimation for these assumptions is given by the fact that the
vehicle movement is triggered by the traffic light, i.e., vehicles are
gathered in front of the them and hence fluctuations can not spread
out. In addition, the cycle length is of the order of the street
length or more precisely, the travel time from one intersection to the
next. It makes no sense to consider cycle times that are much larger
than the travel time which is proportional to the length of the street
segment. Note that the limit $T\to \infty$ corresponds to the case in
which one direction of the network is free to move all the time while
on the other direction it comes to a complete stop. The resulting flow
then is exactly half of the flow found in the underlying NaSch model.

In the following we focus on five scenarios (a)--(e).  The cases (a),
(b) and (c) describe the derivation of the maxima/minima of the
$(v,T)-$curve, (d) gives a calculation of the mean velocity between
maxima and minima, and (e) finally a calculation of the mean velocity
between the minima and maxima. We now discuss these scenarios in more
detail. Note that they are quasi-deterministic and can be slightly
modified in the presence of fluctuations.\\ 

(a) The time a free flowing vehicle requires to move from one intersection
to the succeeding one (one full turn on the periodic street for $N=1$) is
equal to 
\begin{equation}
T_{\text{free}}=\frac{D}{v_{\text{free}}}, 
\label{tfree}
\end{equation}
where $v_{\text{free}}=v_{\text{max}}-p$ is the free-flow velocity
of the underlying NaSch model. In Fig.~\ref{explain2}a a situation 
is displayed where vehicles organize
in a cluster (light grey rectangle) which can move ahead all the
time. This is only possible if the time for one complete traffic
light cycle, i.e., including green  and red phase, is
equal to the cycle time of a vehicle $T_{\text{free}}=T_{\text{green}}
+T_{\text{red}}=2 T$. Obviously this case corresponds to a
maximum in flow whereby the traffic light period is given by
$T= T_{\text{free}}/2$. Additionally there are further maxima when
$T_{\text{free}}=n (T_{\text{green}}+T_{\text{red}})$ with ($n=0,1,2,...$).
Thus the traffic light period 
corresponding to a maximal system flow is given by:
\begin{equation}
T_{\text{max}}=\frac{T_{\text{free}}}{2n}.
\label{casea}
\end{equation}
With similar arguments the occurrence of minima
can be explained. These minima correspond to situations where the
traffic lights switch exactly to red when a vehicle cluster reaches a
intersection. It is clear that the assumptions above are only valid for
very short cycle times $(2T\le T_{\text{free}})$. In the following we will
concentrate on more realistic ``larger'' periods, i.e. 
$2T\geq T_{\text{free}}$.\\

\begin{figure}[!hbt]
\begin{center}
\epsfxsize=0.85\columnwidth\epsfbox{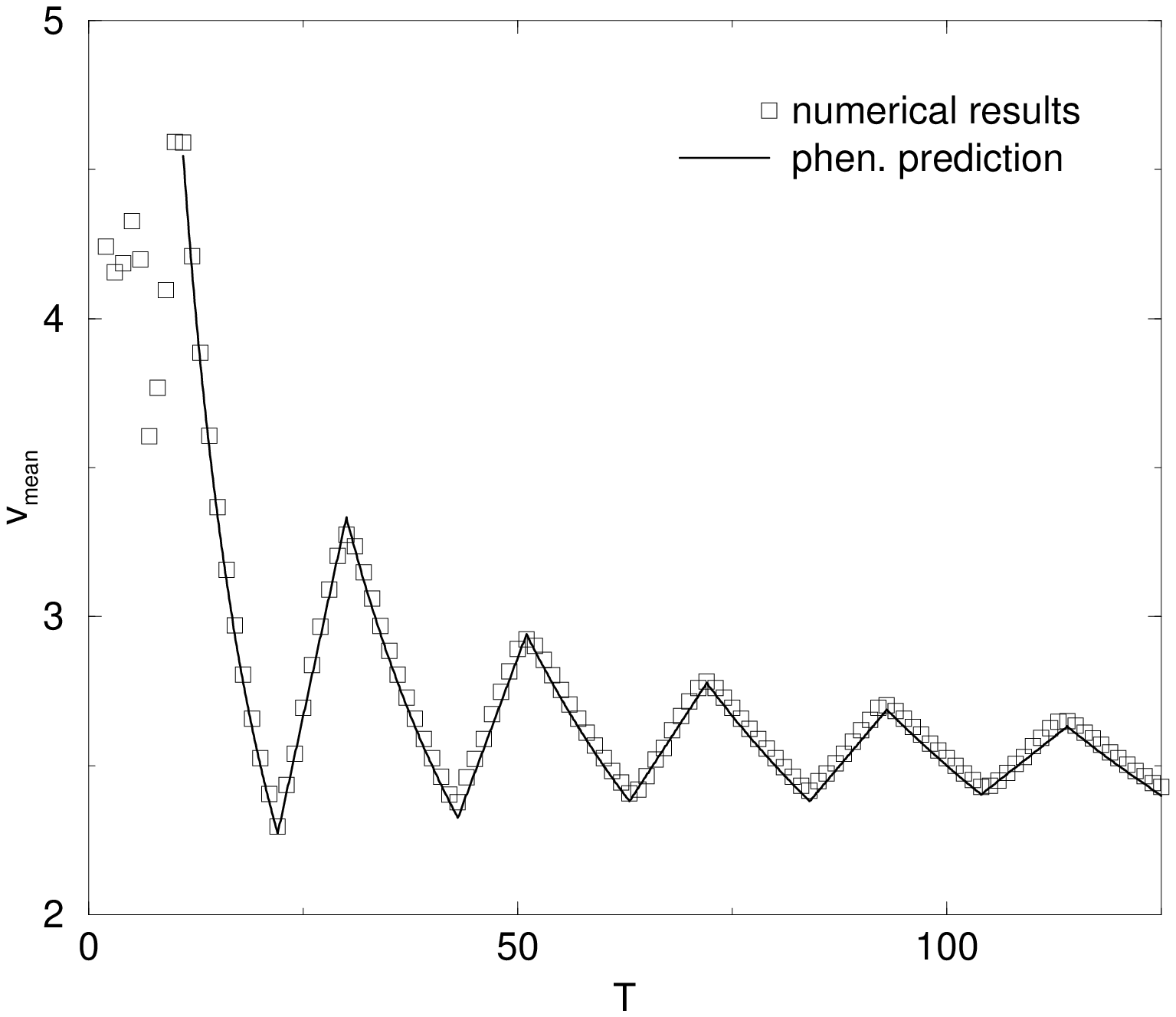} 
\epsfxsize=0.85\columnwidth\epsfbox{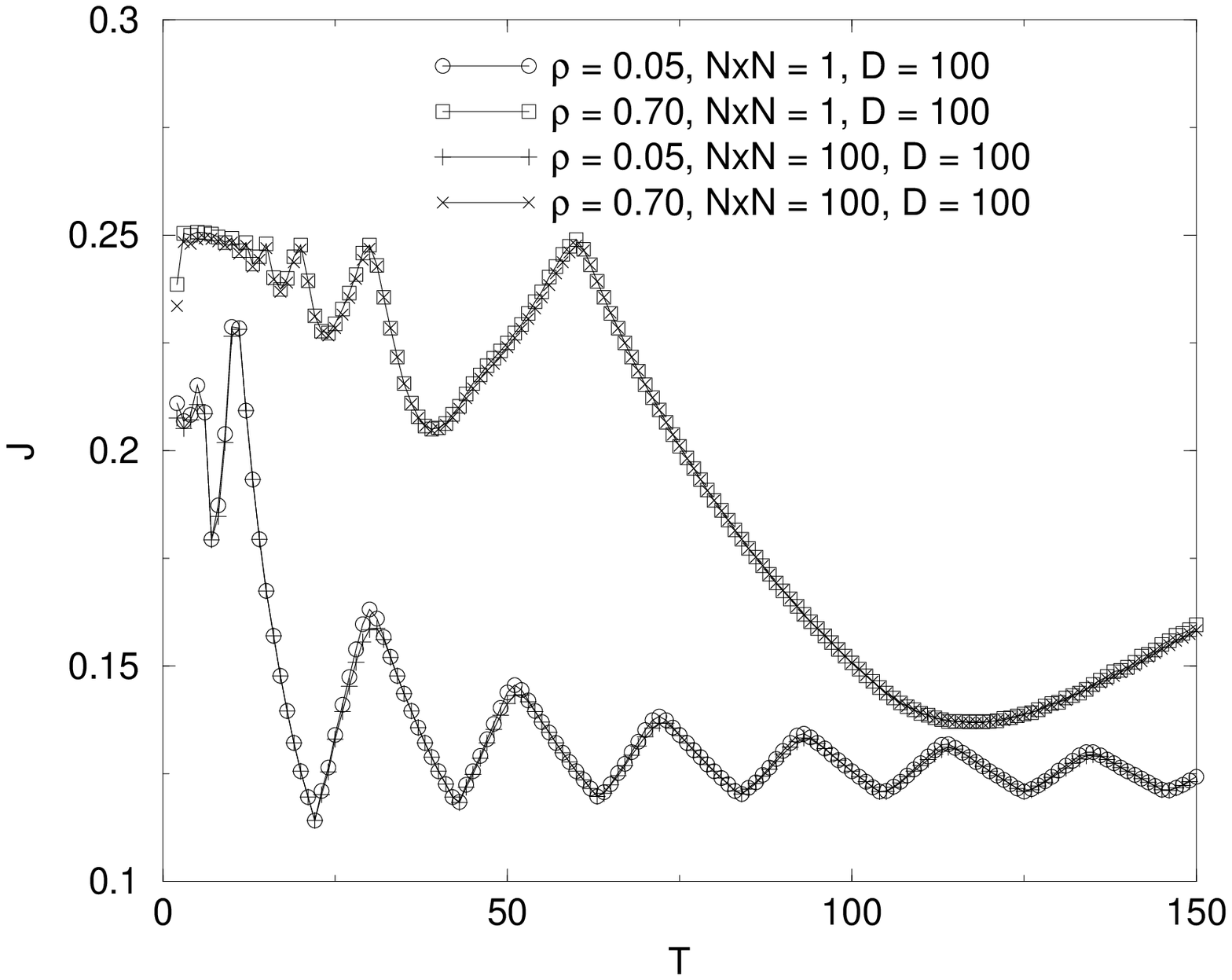}
\end{center}
\vspace{0.2cm}
\caption{{\bf Top:}
The mean velocity $v_{\text{mean}}$ for a minimal network 
$N=1$ is plotted against the cycle time $T$. The
street has a length of $L = 100$ cells and the density is set
to $\rho = 0.05$ (free-flow case). One can clearly see that the
phenomenological approximation agrees very well with the simulation
data. {\bf Bottom:} In order to show, how the small network segment with $N=1$ 
(considered in the heuristic approach) compares to the complete ChSch
city network model we plotted the mean flow against the traffic light
period $T$. This is done once for the ``mini network'' consisting of one
single intersection with a street length of $L = 100$ cells and for a
relatively large network consisting of $N\times N = 100$ intersections 
with $2N^{2}$ street segments each of $D = 100$ cells in length. 
We consider two different densities, one of them corresponding to the 
free-flow density $\rho=0.05$ and the other to a high density state
$\rho=0.7$. Obviously, the deviations in the curves between the large
network and the ``mini network'' are negligible in both density
regimes. The randomization 
parameter is $p = 0.1$ and the maximum velocity is $v_{\text{max}} = 5$ in
both diagrams.}
\label{theo}
\end{figure}

(b) In Fig.~\ref{explain2}b a situation is shown where vehicles
are gathered in front of a red light. After the traffic light switches to
green the vehicles start moving. Then it switches back to
red exactly at the time when the first car of the moving vehicle
cluster reaches the intersection again. Now the complete vehicle
cluster comes to rest and has to wait until the traffic light switches again
to green to continue the movement. Obviously this case corresponds to
a minimum in the flow. The corresponding cycle time is given by the
following assumptions. For this scenario it is sufficient to focus on
the first car of the cluster. At the beginning the first vehicle has
to accelerate to its maximum velocity. This acceleration process will
take on average $T_{\text{acc}}=\frac{v_{\text{max}}}{1-p}$ time-steps. 
After that the vehicle has to trespass the rest of the street until it 
reaches the intersection again. The mean velocity on that part of the road 
is given by $v_{\text{free}}$. The length of this road segment is given 
by the length of the street minus the distance that the vehicle has 
covered during its acceleration phase. Therefore, the time
$T_{\text{first}}=\frac{D-T_{\text{acc}}
(v_{\text{max}}+1)/2}{v_{\text{free}}}$ elapses until
the intersection is reached. In summary, if the chosen cycle time is
equal to
\begin{equation}
T_{\text{min}}=T_{\text{acc}}+T_{\text{first}}+nT_{\text{free}}, 
\label{tmin}
\end{equation}
the system flow is minimal. The last term $n T_{\text{free}}$ (with 
$n=0,1,2...$) takes into account traffic light periods that are larger than the
required time to move from one intersection to the succeeding one or to
make one turn on a periodic system. That way the vehicle cluster is
able to perform $n$ ``turnarounds'' before it has to stop immediately in
front of the ``red light''. These minima at regular distances of
$T_{\text{free}}$ time-steps can be easily identified in
Figs.~\ref{jgegenperiod},~\ref{theo}.\\

(c) In accordance with the occurring minima one can also find maxima
at regular distances (see Figs.~\ref{jgegenperiod},~\ref{theo}). These
maxima correspond to situations where the length of the green time
intervals is sufficiently large so that the last vehicle of a
moving cluster is able to pass the intersection before the traffic
light switches to red. To derive the cycle times corresponding
to this situation one has to focus on the last car. Before the traffic light
switches to green there are $N_{\text{wait}}$ vehicles standing in front 
of it (dark grey rectangle) (see Fig.~\ref{explain2}c). After the
switch to green the last vehicle of the cluster has to wait
on average $T_{\text{wait}}=\frac{N_{\text{wait}}-1}{J_{\text{out}}}$ 
time-steps before the vehicle in front started to move ($J_{\text{out}}$ 
is equal to the flow out of a jam). 
Then further $T_{\text{acc}}$ (see case (b)) time-steps are needed for
the vehicle to accelerate to its maximum velocity. From then on the
vehicle has to reach the first cell (behind the intersection) of the
succeeding street within the remaining ``green light'' interval. The
required time to cover this part of the road is given by
$T_{\text{last}}=\frac{D+N_{\text{wait}}-T_{\text{acc}}
(v_{\text{max}}+1)/2}{v_{\text{free}}}$.
Note, that in comparison to case (b) the last vehicle has to cover a
slightly larger distance than the first one due to its  shifted starting 
position of about $N_{\text{wait}}$ cells. Therefore, the system is in
a state with of maximum flow for the following cycle times:
\begin{equation}
T_{\text{max}}=T_{\text{wait}}+T_{\text{acc}}+T_{\text{last}}
+nT_{\text{free}}. 
\label{tmax}
\end{equation}
As in (b), the last term $n T_{\text{free}}$ takes into account large 
cycles where the vehicle cluster is able to make $n$ full turns
before the pictured situation occurs.\\

(d) We used the previous cases (a)--(c) as a basis for simple heuristic
arguments to derive the cycle times corresponding to maximal
and minimal mean flow states in the system. In the remaining cases we
will show that even the complete dependence of the mean velocity
on the cycle time can be obtained from simple phenomenological
assumptions. For this purpose we focus on a situation where the
vehicle cluster is able to cross the intersection within the ``green
light'', i.e., the traffic light does not switch when the vehicle cluster
occupies the intersection. After the vehicle cluster has passed the
intersection at most $n$ times the vehicles will come to a rest in
front of a ``red light''. The remaining waiting time depends now on
the chosen cycle time. If the traffic light switches to red immediately
before the vehicles reach the intersection the situation corresponds to
minimal flow (see (b)), i.e., the vehicles must wait for the complete
cycle time $T$. Contrary, if the traffic light switches directly after
the cluster has trespassed the intersection the situation corresponds
to the case of maximal flow (see (c)), i.e., the vehicle cluster can
perform a complete turn within a ``red light'' phase and therefore the
remaining waiting time gets minimal. The more general case is given by
a situation between maximal and minimal flow, i.e., the vehicle
cluster is able to pass the intersection and then after a certain
time the traffic light switches to ``red light''. To obtain the mean velocity
of the vehicles within a complete cycle $T_{\text{cycle}}=2 T$ neither
one has to take into account the waiting times of vehicles in the
starting phase nor the acceleration process of the vehicles until the
maximum velocity is reached.
In fact only the driven distance which is equal to $n$ turnarounds
for every vehicle must be considered in order to obtain the mean velocity.
Note, that each vehicle starts its movement out of a certain position
in a waiting queue in front of the traffic light and will occupy exactly the
same position when it comes to a rest again. The mean
velocity is given by
\begin{equation}
\overline{v}_{\text{max--min}}(T,n) = \frac{n D}{2 T}.
\label{vmean}
\end{equation}  
With Eqn.~(\ref{vmean}) it is possible to plot the mean velocity of the
system against the traffic light periods only between each $n$-{\it
th} maximum and $n$-{\it th} minimum of the curve. The shape of the
curve between the $n$-{\it th} minimum and the $(n+1)$-{\it th} maximum
will be discussed in (e). One should keep in mind that the scenarios
(b)--(e) assume $T\ge T_{\text{free}}$.\\

(e) In Fig.~\ref{explain2}e a situation is pictured where the traffic
light switches to ``red light'' within the time interval at which the
vehicle ``cluster'' crosses the intersection. As a consequence the
fraction of vehicles in front of the traffic light will come to a
stop while the rest of the vehicles behind it is able to move
on until they reach the traffic light again (periodic boundary conditions).
The fact that only a fraction of vehicles is able to complete
$n$ cycles whereas other can complete $n+1$ cycles before they are 
forced to stop leads to a simple linear dependence between the mean 
velocity and the cycle time in this area.\\

In the left part of Fig.~\ref{theo} we show how the mean velocity of
the north bound street of the considered ``mini network'' depends on
the cycle time and compare these results with the phenomenological
predictions made in (a)--(e). As one can see the theoretical curve
shows an excellent agreement with the simulation data. Not only the
positions of the maxima and minima are predicted by theory but also
the shape of the curve between the extrema shows a very good agreement
with the numerical results. At this point we want to emphasize that we
checked the mean velocity on the east bound street as well and found
exactly the same results. This is not further surprising if one takes
into consideration that the duration of the traffic light cycles of
both directions are the same, i.e., the time of ``red light'' is equal
to the ``green light'' and when the north-bound direction switches to
green then the east-bound direction switches to red and vice
versa. Therefore the two different directions can be considered as
almost decoupled and independent. Furthermore the right part of
Fig.~\ref{theo} shows that the results obtained from the observed
``mini network'' are completely transferable to large networks. Thus
we stress that the assumptions made in (a)--(e) can be used to adjust
the optimal cycle times in large networks, i.e., in the ChSch model
with synchronized traffic lights. The excellent agreement between the
small and the large network situation can be ascribed to the
synchronized strategy. In fact, there is no difference for a
vehicle approaching an intersection which is a part of a large network
or approaching the only existing intersection due to the periodic
boundary conditions. The state of the traffic lights will be the same
in both cases because of the synchronized strategy. Moreover it is
very interesting that although the vehicle movement is stochastic
(NaSch model) and the mean density on the streets in the network
fluctuates, there is no local concentration of vehicles in the network
leading to remarkable deviations in the flow in comparison to the
idealized ``mini network'' where the density on the streets is fixed.
Note, that this is in contrast to the original formulation of the
ChSch model where a blockage of intersections is allowed. Therefore
fluctuations can lead to a complete breakdown of flow at high
densities where standing vehicles are gathered in one part of the
network. It seems that the signalized intersections of the model
interact with the density fluctuations in a way that the vehicles are
equally distributed in the network. The extreme fluctuations in the
distribution do not play an important role in progress of time because
the blockage of an intersection due to such fluctuations is excluded
here (see sec.~\ref{definition}) and so the density on the roads
fluctuates around a mean value.\\

The results obtained by the phenomenological approach confirm that the
dynamics in the network is driven by the traffic lights and mainly
determined by the distance between them and the density of cars. It
seems that the influence of the model chosen for the vehicle movement
plays a secondary role. We only assume the mean velocity of free
flowing vehicles and the outflow out of a jam as parameters for the
movement from the underlying NaSch model. To verify this, we
investigate a comparable network scenario where the vehicle movement
is realized by the VDR model~\cite{barlo1}. A major difference to the
NaSch model is the occurrence of large phase separated jams and
metastable states in the absence of intersections. However, we found
qualitatively the same results for both models assuming the outflow of
a jam and the mean velocity as parameters. One reason is that the
metastable states of the VDR model are destroyed by disturbances
caused by the traffic lights.\\

So far we have only observed the free-flow case of the ChSch model in
our scenarios. But also for high densities one can find a strong
dependence of the mean flow in the system on the chosen cycle
times (see Fig.~\ref{jgegenperiod}). Obviously for high densities this
dependence is not caused by free flowing vehicle clusters passing or
not an intersection, but rather due to the movement of large jams
gathered in front of the traffic lights. These jams move oppositely to
the driving direction. For densities slightly above the free-flow
density (see $\rho=0.2$ in Fig.~\ref{jgegenperiod}) there are no
characteristic maxima or minima in the mean flow. Here the remaining
jams in the system are small compared to the cycle times, i.e., the
time a jam will block an intersection is negligible small.
Furthermore, for decreasing traffic light cycles, large jams are
divided into smaller ones by the short cycle times. Thus, the mean
flow increases slightly with higher cycle times in this density area
because the number of standing cars decreases. At intermediate
densities (see $\rho=0.5$ in Fig.~\ref{jgegenperiod}) one can find a
similar behavior. As for $\rho=0.2$ the number of jams decreases with
increasing cycle times and the flow grows slightly until it breaks
down at a certain value. This breakdown can be explained as follows:
At high cycle times only one jam remains between two intersections
because the ``red light phase'' is large enough so that all vehicles
are gathered in front of the traffic lights. The breakdown finally occurs when
the ``red light phase'' is even larger than the time needed to
conglomerate all vehicles in front of it. As a consequence,
the vehicles have to wait considerably longer than they are able to
move when further increasing the cycle time. Note, that the motion at
``green light'' is hindered because of the fact that for the
considered densities the jam is relatively large. Therefore a
intersection is blocked when it is reached by the backward moving jam
for a long part of the ``green light phase''. It is interesting that
for high densities (see $\rho=0.7$ in Fig.~\ref{jgegenperiod}) a
strong dependence between the cycle time and the mean flow can be
found with characteristic maxima and minima similar to the free-flow
case. This is caused by the fact that at high densities the dynamics
of the system is completely determined by the movement of a jam. For
example, if the length of one cycle (red light and green light)
is chosen in such a manner that it is equal to the time the downstream
front of a jam needs to move from one intersection to the next one, the
large jam will block the intersection when it is red anyway. This
corresponds to a maximum in the global network flow. The fraction of
time when the ``red-light'' has no influence on the mean flow because
it is blocked by a jam determines the shape of the curve between the
extrema similar to the free-flow scenarios. For a more detailed
discussion, see~\cite{elmar}.  At this point we want to emphasize that
high densities are more difficult to investigate because the jamming
in the NaSch model is strongly determined by the fluctuation
parameter.  For higher $p$ spontaneous jams can occur even in the
outflow region of a jam and therefore jams are not compact anymore.
At high densities one can see a relatively strong influence of $p$
while in the free-flow case the value of the randomization parameter
$p$ does not play an important role.\\
 

\subsection{Green Wave Strategy}

\begin{figure}[!hbt]
\begin{center}
\epsfxsize=0.85\columnwidth\epsfbox{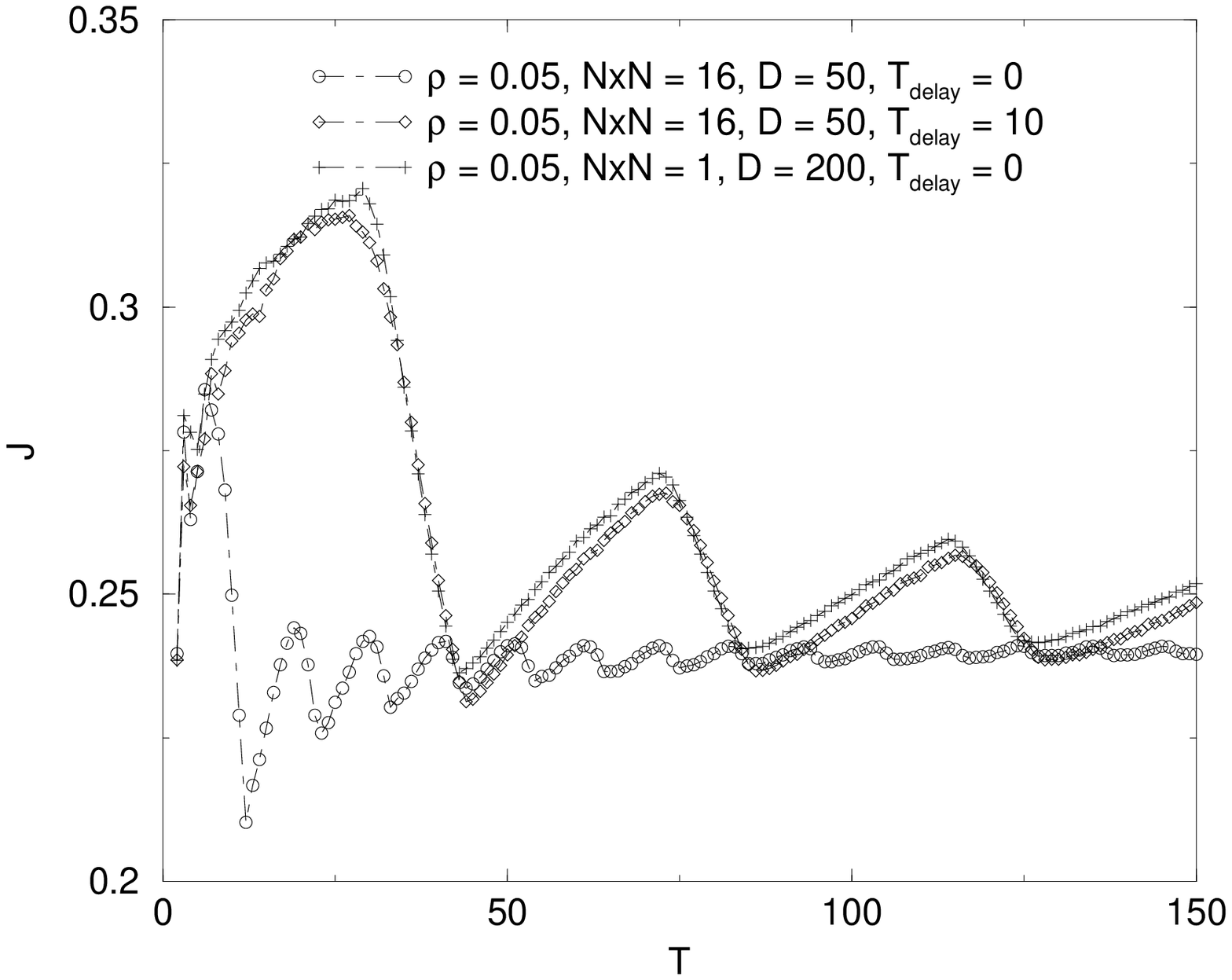}
\epsfxsize=0.85\columnwidth\epsfbox{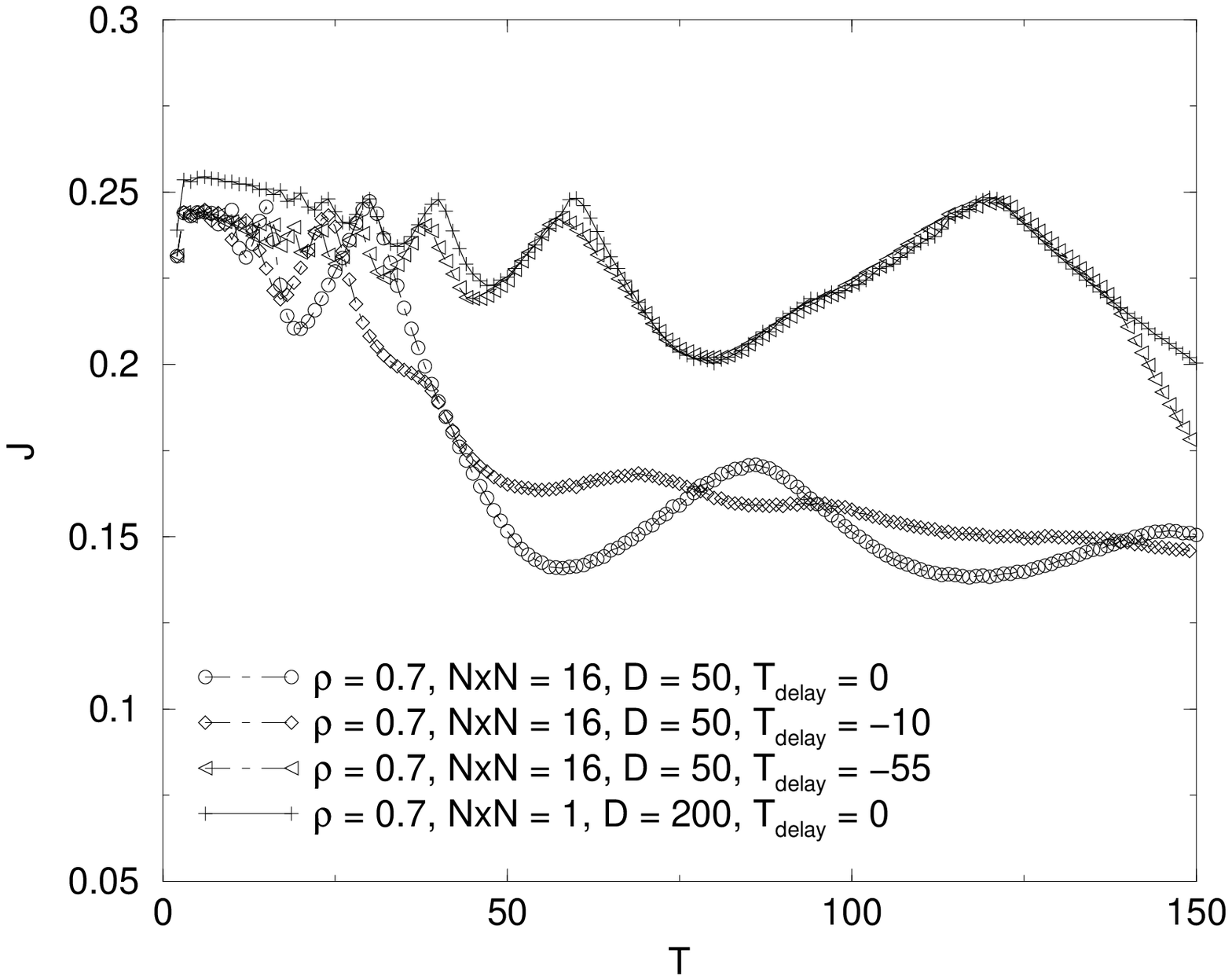} 
\end{center}
\vspace{0.2cm}
\caption{In order
to compare the gain of a network operating with a ``green wave''
strategy to a system with a synchronized strategy we plotted
the flow against the cycle time for both systems. The top
diagram shows the free-flow case of the system. As one can
see, the green wave strategy (time delay $T_{\text{delay}} = 10$)
shows reasonable improvements over the network with synchronized
traffic lights $T_{\text{delay}} = 0$. Moreover for comparing the green wave
strategy with a network consisting of only one intersection, but an equal
total street length, one finds a remarkable agreement.
The bottom diagram shows the influence of the green wave strategy in
the high density state. It its obvious that by definition no green
wave can be established in the system because the density is too high,
so that no jam free states can be obtained. Nonetheless, the
performance of the network with synchronized traffic lights is exceeded by
the ``green wave'' strategy. The randomization 
parameter is $p = 0.1$ and the maximum velocity is $v_{\text{max}} = 5$.}
\label{jgtgreen}
\end{figure}   

In the previous section we discussed the dependence between traffic
light periods and throughput in the ChSch model for synchronized 
traffic lights. It was shown that the whole problem can be reduced to 
an analysis of a single segment (i.e., $N=1$) of the network. This
indicates that synchronizing the traffic lights is an ineffective
strategy which is not capable to bring an additional gain out of the
network topology.
Further it was shown that particularly at free-flow densities there
are strong oscillations in the throughput of the network depending
on the chosen traffic light periods. Another disadvantage is, as one can see
in Fig.~\ref{jgegenperiod}, that the first maxima are located at
unrealistic short cycle times for the chosen street length.\\

In the following we will introduce a simple ``green wave'' 
strategy in order to improve the overall network throughput. Therefore
the ChSch model is enhanced by traffic lights which are not enforced
to switch simultaneously. The intersections are denoted with indices
$i,j$ where $i=0,1,..,N-1$ represents the rows and $j=0,1,..,N-1$ the
columns of the quadratic network. In addition, an individual offset
parameter $\Delta T_{i,j}$ is introduced and assigned to every
intersection. This offset parameter is used to implement a certain
time delay $T_{\text{delay}}$ between the traffic light phases of two
successive intersections. The offset parameter itself can take the values
$\Delta T_{i,j} = 0,...,2T$. Note, that a larger $\Delta T_{i,j}$ has
no effect because $2T$ corresponds to one complete cycle of a traffic
light. The main intention when establishing a ``green wave'' on an
intersected street is to keep a cluster of vehicles in motion. It is
obvious that the optimal strategy is to adjust the time delay between
two successive intersections such that the first vehicle of a moving
cluster trespassing an intersection will arrive at the next traffic
light exactly at the time when it switches to ``green''. This delay is
just the time a free flowing vehicle needs to move from one
intersection to the succeeding one, i.e.,
$T_{\text{free}}=\frac{D}{v_{\text{free}}}$. Thus this is the optimal
time delay $T_{\text{delay}}$ between two intersections. Since we are
interested in constituting the ``green wave'' in the whole network,
two directions must be taken into account. We choose the intersection
at the bottom left corner of the network as the starting point with no
time delay $\Delta T_{0,0} = 0$. Then the offset in the first row will
be chosen as described, i.e., the time delay between two successive
intersections is in the optimal case equal to $T_{\text{free}}$. After
the first row is initialized every intersection in this row will be seen
as a new starting point to initialize the corresponding columns.  In
summary, the offset parameter of the intersections is given by
\begin{equation} 
\Delta T_{i,j} = ((i+j)T_{\text{delay}})\ \text{mod} (2T),\\ (i,j = 0,1,..,N-1),
\end{equation}     
with the optimal offset parameter given by $T_{\text{delay}}=
T_{\text{free}}$, i.e.,
\begin{equation} 
\Delta T_{i,j} = \left((i+j)\frac{D}{v_{\text{free}}}\right) 
\text{mod} (2T),\\ (i,j = 0,1,..,N-1).
\end{equation}     
Using this method a two-dimensional ``green wave'' strategy can be
established in the ChSch model.\\

To quantify the improvement obtained by the ``green wave'' strategy the
overall network flow is plotted against the cycle time (see
Fig.~\ref{jgtgreen}) and compared with the synchronized strategy.
The left diagram corresponds to the free-flow case of the
system. The density is chosen to $\rho=0.05$ to ensure
that moving vehicles are able to drive from one intersection to the next
one without being constricted by standing cars. Obviously, the green
wave strategy with a properly chosen offset parameter, i.e., for the
considered street length equal to $T_{\text{free}} = T_{\text{delay}} = 10$,
shows reasonable improvements over the strategy with synchronized
traffic lights ($T_{\text{delay}} = 0, N = 4$). The whole spectrum of
plotted cycle times $T$ for the ``green wave'' strategy exceeds
the performance of the network with synchronized traffic lights or at least
keeps the performance. Moreover, comparing the green wave strategy
to a network consisting of only one intersection, but with the same total
street length, one finds a remarkable agreement of the curves. Note, that 
every street in the considered network with $N = 4$ is intersected
four times. We want to stress here that for free-flow densities
in the ChSch model the ``green wave'' strategy is capable to pipe all
vehicles through the streets, i.e., for the vehicles on the streets it
seems as if there is only one intersection in the system left due to the fact
that the remaining ones are always green when approached by the
vehicle cluster. Further we want to point out that similar to the
case with a synchronized strategy the traffic lights interact with the
vehicles in such a way that a ``green wave'' is established in the
network independent of the initial vehicle distribution or the density
fluctuations caused by the internal stochasticity of the model. 
Recapitulating, one of the most important benefits of the green wave strategy
is the fact that a street with total length $L$ consisting of $N$
street segments, each with a length $D$, behaves like a street intersected 
only once (see Fig.~\ref{jgtgreen}). Therefore the optimal
cycle time of a traffic light corresponding to the maximal flow is shifted 
towards realistic values (see Sec.~\ref{synchrostrat}a) even for small 
street segment lengths $D$. One obtains the following equation for the 
cycle time corresponding to maximal flow (see Eqn.~\ref{casea}): 
\begin{equation}
T_{\text{max}} = \frac{L}{2v_{\text{free}}} = \frac{ND}{2v_{\text{free}}}.
\end{equation}
\\

As one can see in the right part of Fig.~\ref{jgtgreen} even for high
densities the ``green wave'' strategy shows an incisive impact to the
network flow. Although by definition no ``green wave'' can be
established at high densities (for the chosen density of $\rho = 0.7$
no jam free state can exist), an offset in the switching between
successive traffic lights can lead anyhow to an improved flow. The origin of
this improvement is completely different in comparison to the free-flow 
case. For low densities the dynamics is driven by vehicles
organized in clusters which can move through the streets undisturbed 
due to the optimal strategy whereas the dynamics for high
densities is governed by the motion of large jams. Large jams move
oppositely to the driving direction of the vehicles from one
intersection to the one before. Due to their spatial extension a intersection
is blocked for a certain time when trespassed by a jam. Thus the
optimal system state would be reached if a jam moves backward from one
intersection to the one before and blocks it while the
traffic light is red anyway so that afterwards moving vehicles (outflow of
the jam) can take advantage of the green phase
as much as possible. In fact, the portion of time 
that a intersection is blocked or free determines
the system flow. 
Note, that the time delay at high densities has to be
negative since jams move opposite to the driving direction.
For a time delay in the order of the optimal time delay of the free-flow 
case (see Fig.~\ref{jgtgreen} (right) for  $T_{\text{delay}} = -10$) the 
curves corresponding to the ``green wave'' strategy and the synchronized
traffic lights do not differ much because this $T_{\text{delay}}$ is determined
by the free vehicle movement. Considering instead the velocity of a
jam which is approximately about $v_{\text{jam}}=\frac{1}{1-p}$
(see \cite{neubert}) and assuming that the optimal time delay is
the travel time $T_{\text{jam}}=\frac{D}{v_{\text{jam}}}$ for the backward 
motion of a jam between two intersections, the difference to the synchronized
case gets transparent (see Fig.~\ref{jgtgreen} (right) for
$T_{\text{delay}} = -55$). The ``green wave'' strategy allows now a 
reasonable improvement over the synchronized strategy.
Similar to the free-flow density case, the performance of the network
with synchronized strategy is exceeded by the ``green wave'' strategy
for almost all cycle times. Moreover, comparing the ``green
wave'' strategy with an optimal time delay to an idealized ``mini
network'' consisting of only one intersection, but with an equal total
street length one finds an reasonable agreement between the curves as well. 
This indicates that for high densities jams can be guided perfectly
through the streets by a ``green wave'' strategy. However, one has to
recognize that strong oscillations at high densities depend on the
statistics of the underlying NaSch model so that the expected gain at
these high densities will decrease with increasing $p$.
\label{green}  


\subsection{Random Offset Strategy}

\begin{figure}[!hbt]
\begin{center}
\epsfxsize=0.85\columnwidth\epsfbox{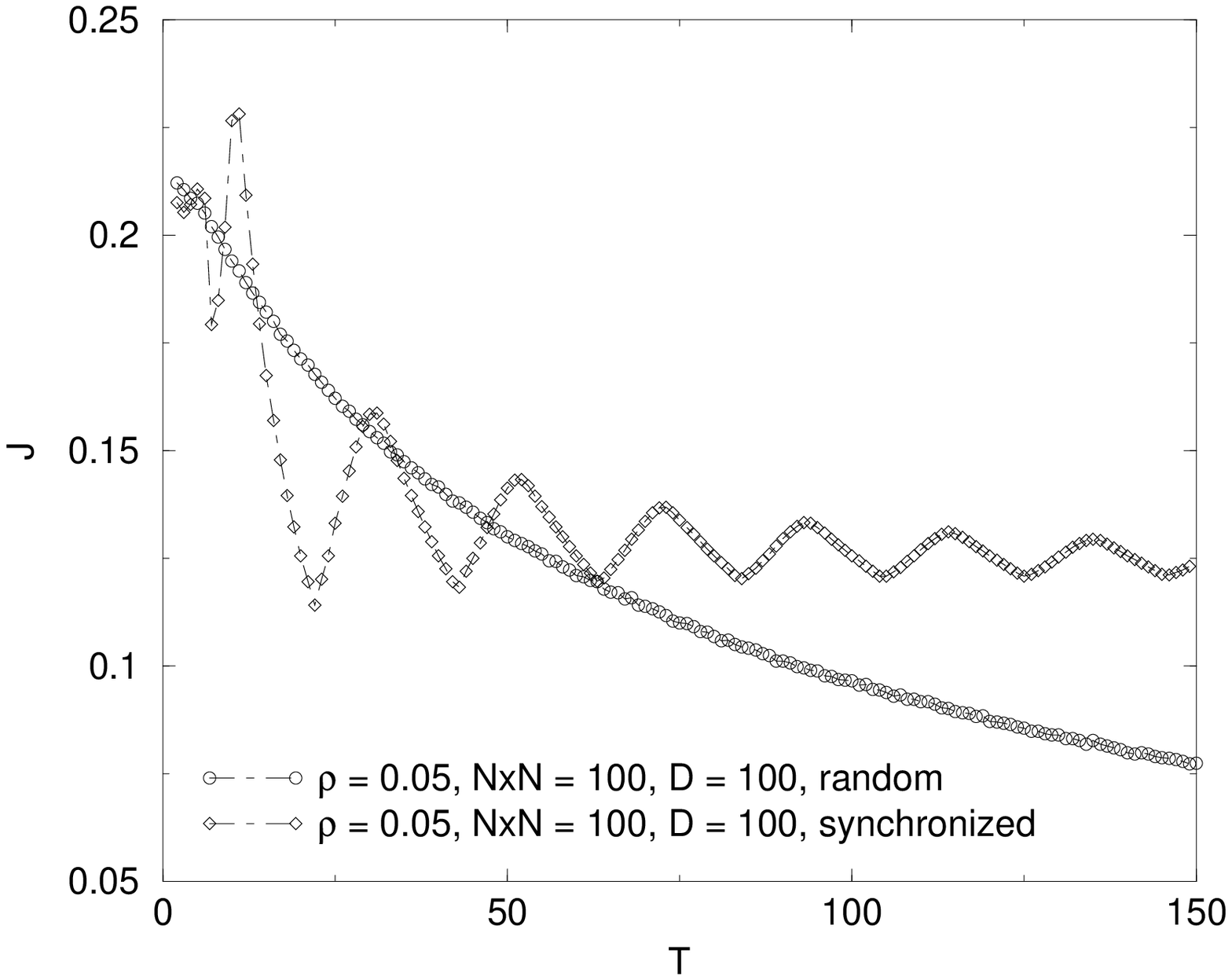}
\epsfxsize=0.85\columnwidth\epsfbox{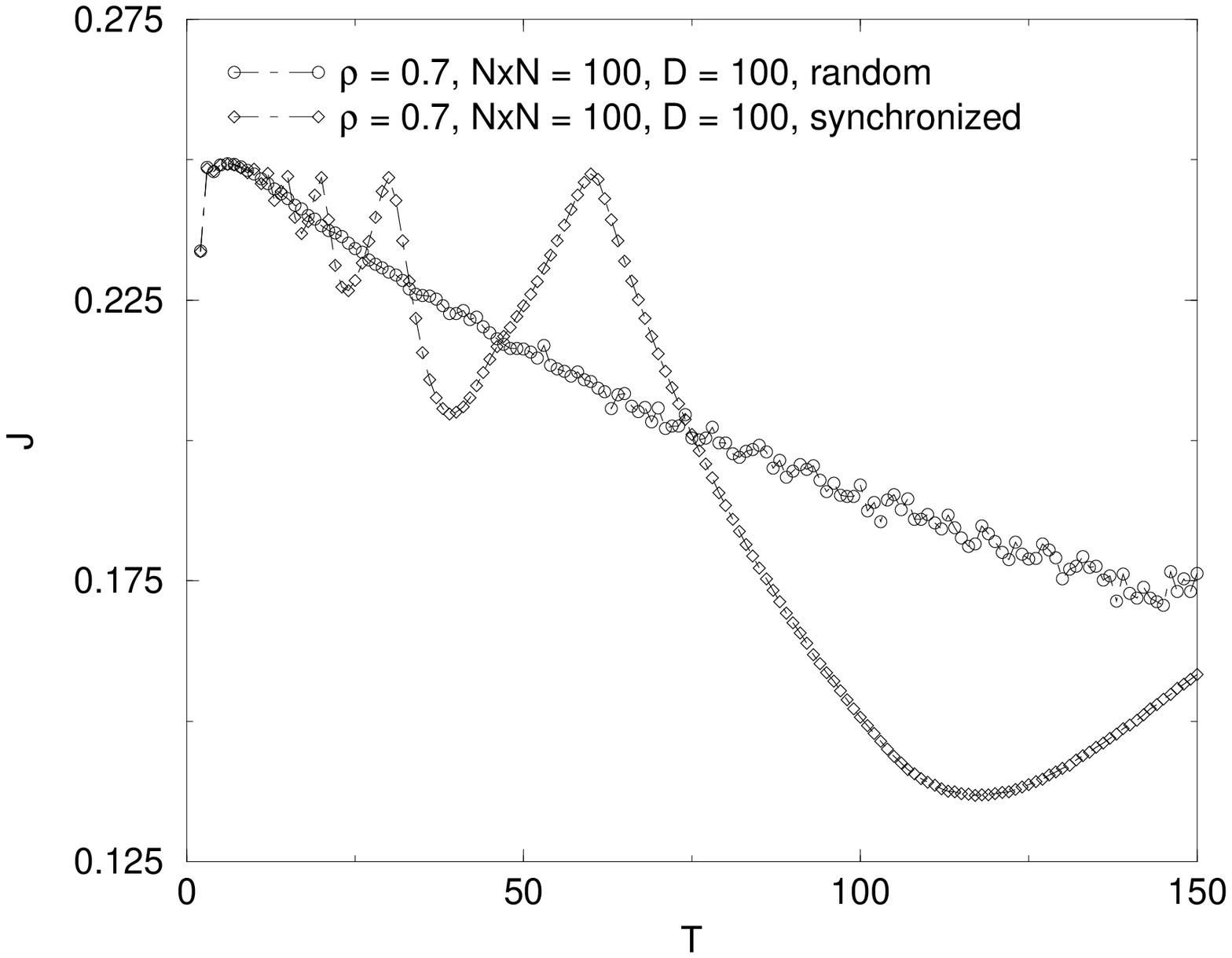}
\end{center}
\vspace{0.2cm}
\caption{The random
offset strategy is compared to the original ChSch model with
synchronized traffic lights. The mean flow is plotted versus the
traffic light periods for the two different strategies. The 
network consists of $N\times N = 100$ intersections with
$2N^2$ street segments each of length $D = 100$ cells. 
{\bf Top:} In the left part of the
Fig. we chose a low density (free-flow regime, $\rho = 0.05$). It
can be seen clearly that the oscillations found in the synchronized network
are suppressed by the random offset strategy. Furthermore in the free
flow density regime the random offset strategy shows some advantages
over the synchronized strategy, but only for low cycle times.
{\bf Bottom:} The oscillations for high densities ($\rho = 0.70$) are
suppressed in a similar manner as for the low density case. In
addition, the random offset strategy seems to outperform the
synchronized strategy in the whole plotted area. The randomization 
parameter is $p = 0.1$ and the maximum velocity is $v_{\text{max}} = 5$.}
\label{jgtrandom}
\end{figure}

In this section we want to point out that switching successive
traffic lights with a random shift instead of a fixed time delay can lead to
a more flexible strategy, e.g., without oscillations. Moreover it 
will be shown that in contrast to a system with synchronized traffic lights a random
shift between the intersections can lead to a remarkable higher global
system flow. As in the previous section the traffic lights are not
enforced to switch simultaneously anymore. For this purpose an
individual offset parameter $\Delta T_{i,j}$ is introduced and
assigned to every intersection (see previous section for a detailed
explanation). The offset parameter itself can take values between
$\Delta T_{i,j} = 0,...,2T$ which are chosen in the following
from an equally distributed random distribution.\\ 

To give an insight into the effects induced by random offsets we
depicted the throughput in the network in dependence of the 
cycle times in Fig.~\ref{jgtrandom}. The random offset strategy is
compared to the ChSch model with synchronized
strategy. Obviously the strong oscillations found in the
curves corresponding to the synchronized strategy are destroyed by the
randomness in the switching. Thus the random offset strategy
leads to a smoothed curve which is very useful adjusting the
optimal cycle times in a network. One is no longer forced to pay strong
attention to the cycle times like in systems with synchronized
or ``green wave'' strategies.\\
 
The left part of the Fig.~\ref{jgtrandom} shows a system with
free-flow density $\rho = 0.05$. The random offset strategy
outperforms the synchronized strategy only for relatively low 
cycle times because unfavorable states (states with minimal global
flow) are avoided by the randomness. For higher cycle times the
global flow in a system with random offset strategy falls clearly
below the global flow in a system with synchronized
strategy. In the case of a system with synchronized traffic lights the curve
converges in the limit $T\to \infty$ to the half of the flow
found in the NaSch model. This corresponds to the case in which
vehicles in the network are free to move in one direction all the time
while in the other direction it comes to a complete stop. In contrast,
the flow in the random offset strategy converges to zero since the switching is
not synchronous and therefore the traffic lights along one direction are 
green or red at random so that all vehicles are gathered in front of the
red lights. Additionally, one has to consider that although the random
offset strategy is very effective for low cycle times one can obtain
higher flows with the ``green wave'' strategy.\\

At high densities ($\rho = 0.70$ in Fig.~\ref{jgtrandom}), the oscillations 
are suppressed in
a similar manner as for the low density case. Hence, as for low
densities, this strategy gives an improved flexibility when adjusting
optimal cycle times in the network. In addition, the random
offset strategy outperforms the synchronized strategy not only for low
cycle times, but also in the whole range plotted in 
Fig.~\ref{jgtrandom} except for some peaks.
One obvious explanation for the profit out of the randomly switching
traffic lights is that parts of the network are completely jammed while in
other parts of the network the cars can move nearly undisturbed.
However, the flow obtained by the ``green wave'' strategy is still
remarkably higher than the flow obtained by the
random offset strategy. Furthermore one has to consider that the strong
oscillations at high densities depend on the statistics of the
underlying NaSch model so that the expected gain at this high
densities will decrease with increasing randomization parameter $p$. 
Thus we want to point out that among the analyzed global strategies 
the ``green wave'' strategy leads to the highest global flow in the 
network for free-flow densities as well as for high density states 
while the ``random offset'' strategy provides the greatest flexibility
hence the oszillations are suppressed.\\


\section{Summary and Discussion}

We have analyzed the ChSch model which combines basic ideas from the
Biham-Middleton-Levine (BML) model of city traffic and the Nagel
Schreckenberg (NaSch) model of highway traffic. In our investigation
we focused on global traffic light control strategies and tried to
find optimal model parameters in order to maximize the network flow. For
this purpose we started with the original formulation of the ChSch model
where the traffic lights are switched synchronously. It is shown that
the global throughput of the network strongly depends on the 
cycle times, i.e, one finds strong oscillations in the global flow in
dependence of the cycle times both for low as well as for high
densities. A simple phenomenological approach has been suggested for
the free-flow regime in order to determine the
characteristics in regard to the model parameters and to obtain a
deeper insight into the dynamics in the network. The phenomenological
results show a very good agreement to numerical data and indicate
that the choice of the underlying model for vehicle movement between 
intersections does not play an important role. Thus we want to stress 
here that the global throughput in the ChSch model is mainly determined 
by the travel times between intersections which depends on the
length of the street segments and the density and maximal velocity
of the cars.\\

In order to allow a more flexible traffic light control the ChSch model was
enhanced by an additional model parameter. This new parameter is
assigned to every intersection representing a time offset, so that the
traffic lights are not enforced to switch simultaneously anymore. A
two dimensional ``green wave'' is implemented with the help of the new
parameter. The ``green wave'' gives much improvement to the flow in 
comparison to the synchronized strategy at low densities and has even an 
incisive impact on the throughput at high
densities. Moreover it is shown that the influence of intersections
along a street is completely avoided by the ``green wave'' strategy
because the results can be compared with results obtained from a
system containing only one single intersection instead of many
others. Although the ``green wave'' strategy is capable to give a strong
improvement, the dependence between flow and the cycle time
found in the original ChSch model remains. Thus to avoid this strong
oscillations we further analyzed a network where traffic lights are
switched at random. It is shown that the strong oscillations found for
a synchronized strategy and for the ``green wave'' strategy are
completely suppressed by randomness. Thus the random offset strategy
can be very useful if a control strategy is required which is not very
sensitive to the adjustment of the cycle
times. Moreover, the random offset strategy outperforms the standard
ChSch model with synchronized traffic lights at low densities for small cycle
times and at high densities for all cycle times.
An explanation for the profit at high densities is the
fact that some parts of the network are completely jammed while in
other parts of the network the cars can move nearly undisturbed. This
additional gain due to the inhomogeneous allocation of vehicles
indicates that an autonomous traffic light control based on local
decisions could be more effective than the analyzed global 
shemes. In~\cite{huberman} Faieta and Huberman investigated an autonomous
traffic light strategy which shows a very good performance. 
Results of simulations with the ChSch model about the impact
of traffic lights which are autonomously adapted to the traffic
conditions by suitable parameters will be presented in~\cite{city2}.\\

To conclude, the results presented here are of practical relevance for
various applications of city traffic. Due
to its simplicity cellular automata models have become quite popular
for large scale computer simulations whereby especially city traffic
with its complex network topology is one of the favorable
applications. In particular the knowledge of the impact of topological
factors in regard to certain traffic control strategies can be
very benefitable when studying various kinds of city networks, even
those with a more sophisticated topology than those implemented in the
ChSch model.\\     
 
{\em Acknowledgement:}\\
We thank Torsten Huisinga, Wolfgang Knospe, and Andreas Pottmeier
for useful discussions.


\end{document}